\documentclass[aps,reprint,prb,showpacs,amsmath,floatfix,superscriptaddress]{revtex4-1}

\usepackage{graphicx}
\usepackage{bbold}
\usepackage{times}
\usepackage{hyperref}
\usepackage{dsfont}
\usepackage{color}
\begin{document}

\title{Phase boundary location with information-theoretic entropy in tensor renormalization group flows}

\author{Adil A. Gangat}
\affiliation{Department of Physics, National Taiwan University, Taipei 10607, Taiwan}
\affiliation{Département de Physique, Université de Sherbrooke, Sherbrooke, Québec, Canada J1K 2R1}
\affiliation{School of Physics, Georgia Institute of Technology, Atlanta, GA 30332, USA}
\author{Ying-Jer Kao}
\email{yjkao@phys.ntu.edu.tw}
\affiliation{Department of Physics, National Taiwan University, Taipei 10607, Taiwan}
\affiliation{National Center for Theoretical Sciences, National Tsing Hua University, Hsinchu 30013, Taiwan}
\affiliation{Department of Physics, Boston University, Boston, Massachusetts 02215, USA}
\date{\today}
\begin{abstract}
We present a simple and efficient tensor network method to accurately locate phase boundaries of two-dimensional classical lattice models.  The method utilizes only the information-theoretic (von Neumann) entropy of quantities that automatically arise along tensor renormalization group [Phys. Rev. Lett. \textbf{12}, 120601 (2007)] flows of partition functions.  We benchmark the method against theoretically known results for the square-lattice $q$-state Potts models, which includes first-order, weakly first-order, and continuous phase transitions, and find good agreement in all cases.  We also compare against previous Monte Carlo results for the frustrated square lattice $J_1-J_2$ Ising model and find good agreement.
\end{abstract}

\maketitle

\section{Introduction}

Tensor networks serve as a powerful ans\"{a}tze for many-body quantum wavefunctions and many-body classical partition functions \cite{cirac2009renormalization, orus2014practical}.  Algorithms to find tensor network representations of quantum wavefunctions or classical partition functions entail discarding the irrelevant portions of Hilbert space or state space to find representations that are numerically efficient.  The pioneering example of this is the density matrix renormalization group (DMRG) \cite{white1992density}, devised for wavefunctions of one-dimensional quantum lattices, but also applicable to two-dimensional classical lattices \cite{nishino1995density} through the well known quantum-classical correspondence.  The first developments explicitly for two-dimensional classical lattices were transfer matrix-based algorithms \cite{nishino1996,nishino1997corner,murg2005efficient}, followed by the block spin-like tensor renormalization group (TRG) \cite{levin2007tensor} algorithm.  Variations of TRG were then developed \cite{xie2009second,gu2009tensor,zhao2010renormalization,xie2012coarse,evenbly2015tensor,yang2017loop,bal2017renormalization,morita2018tensor,hauru2018renormalization,evenbly2018gauge,harada2018entanglement,nakamura2019tensor,iino2019,adachi2019} for better accuracy and enhanced capability, such as applicability in higher dimensions .  These tensor network algorithms do not suffer from the notorious sign problem that sometimes arises in Monte Carlo.

Tensor network methods can be used to locate and characterize phase boundaries of classical lattice models either via the type of thermodynamic analysis that is done with Monte Carlo or through non-thermodynamic analysis.  Thermodynamic analysis entails calculation of higher order moments of physical quantities, which is a highly non-trivial task with tensor networks.  Nevertheless, a recent work \cite{morita2019calculation} showed how to accomplish this to high accuracy with higher order TRG (HOTRG) \cite{xie2012coarse} and used it to compute the phase transition temperature, transition order (i.e. first-order vs. continuous), and critical exponents for the $2$-,$3$-,$4$-,$5$-, and $7$-state Potts models on the square lattice. Non-thermodynamic analysis can be done through computations with the fixed point tensors from TRG-type algorithms.  These fixed point tensors encode the degeneracy of the phase in a very simple way \cite{gu2009tensor}, and an abrupt change in the degeneracy indicates a phase transition. This approach has been used, for example, with HOTRG to compute the critical temperature of the 2-state Potts model on the simple cubic lattice to very high precision \cite{shun2014phase}.  Additionally, computations of the central charge and scaling dimensions from the fixed point tensors \cite{gu2009tensor,evenbly2015tensor,hauru2018renormalization} can locate continuous phase transitions and yield their critical exponents (exceptional cases may be continuous transitions that do not have conformal invariance, such as the phase transition in the $4$-state Potts model on the square lattice).

In this work we deal with an alternative non-thermodynamic quantity for phase boundary location: von Neumann entropy.  A few recent works \cite{krvcmar2015reentrant,krcmar2016phase,krvcmar2016phase,huang2017holographic} have utilized von Neumann entropy to locate phase boundaries of 2d classical lattices with the corner transfer matrix renormalization group (CTMRG) algorithm \cite{nishino1996,nishino1997corner}, but we are not aware of any works that do so with TRG.  Here we explain the straightforward use of von Neumann entropy for phase boundary location of 2d classical lattice models with the TRG algorithm.  In contrast to the thermodynamic approach, the von Neumann entropy TRG method presented here is vastly simpler because it does not require computation of higher order moments.  In contrast to the phase degeneracy method, the von Neumann entropy TRG method does not require coarse graining deep into the thermodynamic limit and does not require encoding of symmetries (numerical instability can blur the transition point in the phase degeneracy method if symmetries are not encoded, as seen in Ref.~\onlinecite{yang2016tensor}).  In contrast to the approach of computing central charge, which locates only continuous transitions, the von Neumann entropy TRG method can locate both first-order and continuous transitions.

On the other hand, the von Neumann entropy TRG method is specific to only two-dimensional lattices and can not characterize the phase boundary, whereas the other approaches (both thermodynamic and non-thermodynamic) are applicable in higher dimensions as well and can characterize the phase boundary (except for the phase degeneracy method).

The use of von Neumann entropy as a signal for 2d classical phase transitions comes through the well known correspondence between (1+1)d quantum and 2d classical models. In particular, the phase transition of a 2d classical model coincides with a phase transition of the same type in a corresponding (1+1)d quantum model.  In a (1+1)d quantum model the von Neumann entropy of the reduced density matrix of a (sufficiently large) contiguous subsystem is maximal at a phase transition, and this entropy maximum also marks a phase transition in the corresponding 2d classical model.  TRG simulations of 2d classical models have direct access to the von Neumann entropy of the corresponding (1+1)d quantum system, and can therefore use it to locate the phase boundaries of 2d classical models. We show below that tuning 2d classical models to maximize this von Neumann entropy in TRG simulations gives the location of their phase transitions to good accuracy.

In the following sections we first review the relevant theoretical background of TRG (sections II. and III.), then describe the implementation of our method (section IV.).  In section V. we benchmark against the theoretically known transition temperatures of the square lattice q-state Potts models, which exhibit different transitions (depending on the value of $q$): first-order, weakly first-order, and continuous. In section VI. we apply our method to the frustrated $J_1-J_2$ Ising model on the square lattice and compare our results to previously published Monte Carlo results.

\section{TRG flows of partition functions near phase boundaries}
Partition functions of two-dimensional classical lattices can be represented as contractions of two-dimensional networks of tensors \cite{levin2007tensor} where each tensor corresponds to a few lattice sites.  An example given in Ref.~\onlinecite{levin2007tensor} for the partition function ($Z$) of a honeycomb lattice model is
\begin{align}
Z=\sum_{ijk...}A_{ijk}A_{ilm}A_{jnp}A_{kqr}...,
\label{partitionfunction}
\end{align}
where $A_{ijk}$ is a three-leg tensor corresponding to three microscopic degrees of freedom.  The TRG algorithm begins with a few (or even just one) tensors at the UV scale and applies a succession of steps, each of which simultaneously grows and coarse grains the lattice.  The growth of the lattice is exponential in the number of TRG steps, which makes calculation of the thermodynamic partition function computationally feasible: after tens of TRG steps, a single tensor represents many degrees of freedom rather than just a few, and tracing over only one or a few tensor(s) becomes sufficient to approximate the partition function in the thermodynamic limit.  For example, in the case of a monopartite square lattice the TRG algorithm may start with a single tensor corresponding to a system size of $2\times2$, and after $N$ TRG steps end with a single tensor that corresponds to a system size of $2^{N/2} \times 2^{N/2}$.

TRG coarse graining entails an information compression scheme, based on the singular value decomposition, that in many cases allows the coarse grained tensors of a partition function to have low compression error (a.k.a. ``truncation error") while still keeping the dimension of their indices (a.k.a. ``bond dimension") within computationally feasible limits.  More precisely, the minimum bond dimension required for maintaining low truncation error grows with each coarse graining step in the early part of the TRG flow of a partition function, but saturates to a finite value when the coarse graining approaches the correlation length of the system.  Near criticality, however, the correlation length diverges, and TRG breaks down in the sense that the minimum bond dimension required for maintaining low truncation error grows without saturating at a computationally feasible value. TRG coarse graining to the thermodynamic limit with low truncation error therefore becomes computationally prohibitive for partition functions near criticality (i.e. the tensors required become too large).  Similarly, the finite but very large correlation lengths that are a hallmark of \textit{weakly} first-order phase transitions can also make low-loss TRG flows to the thermodynamic limit computationally prohibitive.  Crucially, \textit{our method of using TRG flows to locate phase boundaries does not require the TRG flows to always maintain low truncation error}. Therefore, TRG flows with bond dimension ($\chi$) fixed at computationally modest sizes are sufficient for our method.

\section{TRG von Neumann entropy as ground state entanglement entropy}
The correspondence between 2d classical and (1+1)d quantum systems means that a theoretical understanding of our method can be gained by considering the entanglement properties of ground states of (1+1)d quantum spin chains and their finite-$\chi$ tensor network representations.  We discuss here the specific case of matrix product states (MPSs) \cite{fannes1992finitely, ostlund1995thermodynamic,perez608197quantum,vidal2007classical,orus2008infinite} due to the availability of relevant results. The wavefunction of a  (1+1)d quantum spin chain with $N$ sites may be represented as a MPS: $|\Psi\rangle=\sum_{s_1,...,s_N=1}^d \textrm{Tr}(A_1^{s_1}...A_N^{s_N})|s_1...s_N\rangle$, where the $A_j$ are tensors of dimension $d \times \chi \times \chi$, $d$ is the dimension of spin $s_j$ at site $j$, and $\chi$ is again referred to as the ``bond dimension".

For a bipartite quantum system $AB$ in a pure state, the subsystems $A$ and $B$ may each still have a mixed (i.e. uncertain) state due to quantum correlations (i.e. entanglement) between $A$ and $B$.  The resulting ``entanglement entropy" of subsystem $A$ is defined as $S=-\textrm{tr}(\rho_A \textrm{log}_2 \rho_A)$.  It is useful to consider the entanglement entropy of a contiguous subblock in both infinite and finite (1+1)d quantum spin chains.  In the ground state of the chain, the entanglement entropy of such a contiguous subblock, as well as its MPS representation, exhibits universal properties near and at criticality \cite{amico2008entanglement, holzhey1994geometric, osterloh2002scaling, vidal2003entanglement, latorre2003ground, calabrese2004entanglement, refael2004entanglement, tagliacozzo2008scaling, pollmann2009theory, pirvu2012matrix}.

In the ground state of an infinite chain, the entanglement entropy of the infinite half-chain diverges near criticality as $S\propto\textrm{log}(\xi)$, where $\xi$ is the correlation length of the system.  In an infinite MPS (iMPS) \cite{vidal2007classical, orus2008infinite} representation, however, the finite bond dimension causes the entanglement entropy to saturate to a finite maximum near the critical point (the distance from the critical point goes to zero as $\chi\rightarrow\infty$) \cite{tagliacozzo2008scaling, pollmann2009theory}.  The finite $\chi$ of iMPSs also leads to a $\chi$-dependent universal scaling behavior of local observables, which enables a ``finite entanglement scaling" (FES) \cite{tagliacozzo2008scaling, pollmann2009theory} analysis analogous to the well known finite size scaling (FSS) analysis.

A finite-length ($L$) contiguous subblock of an infinite chain has entanglement entropy constant in $L$ off criticality and logarithmic in $L$ at criticality (for sufficiently large $L$) \cite{vidal2003entanglement}.  This behavior also occurs for contiguous subblocks of sufficiently large \textit{finite} chains in the ground state\cite{vidal2003entanglement}, but with a modification: at criticality the entanglement entropy grows logarithmically over a finite range of $L$ but then saturates and starts to decrease after reaching half the chain length due to the finiteness of the system.  Thus, in either case (infinite chain or large, finite chain), there is a range of $L$ for which the entanglement entropy is maximal at the critical point.  For finite chains, MPS representations with sufficiently large $\chi$ can reproduce this behavior.  Further, it was shown for the case of periodic boundary conditions that such finite MPSs exhibit a crossover (as a function of $\chi$ and system size) between regimes where either FSS or FES is valid \cite{pirvu2012matrix}.  For the entanglement entropy this means a crossover between $S\propto\textrm{log}(L)$ and $S\propto\textrm{log}(\chi)$.

The upshot is that in all of the above cases of (1+1)d quantum systems and their finite-entanglement (i.e. finite-$\chi$) tensor network representations, the critical point can be approximately identified with the parameter value that maximizes the entanglement entropy.  The same must also be true for first order transitions if the correlation length on both sides of the transition is maximal at the transition (this is intuitively expected to be the usual case). In the present work, we wish to investigate how well this information-theoretic way of locating phase boundaries of (1+1)d quantum systems translates to two-dimensional classical lattice models via the quantum-classical correspondence in TRG.

The quantum-classical correspondence in the case of TRG is such that the tensor at each step in a TRG flow of a classical 2d partition function corresponds to a representation of a (1+1)d quantum ground state imagined to live on the boundary of the classical system \cite{levin2007tensor}.  The gap of the corresponding classical and quantum systems is the same, and each tensor leg corresponds to a contiguous subblock of the periodic (1+1)d quantum system whose size grows exponentially in the number of TRG steps.  For the example of a square lattice, this can be summarized as
\begin{equation}
|\Psi\rangle\approx \sum_{ijkl}A_{ijkl} |\psi_i\rangle|\psi_j\rangle|\psi_k\rangle|\psi_l\rangle,
\end{equation}
where $|\Psi\rangle$ is the boundary ground state wavefunction, $A_{ijkl}$ is the TRG tensor, and $|\psi_i\rangle$ is a pure state of the contiguous subblock corresponding to tensor leg $i$.  
Here it is manifest that the TRG tensor encodes the entanglement between the subblocks of the (1+1)d quantum chain.  At each TRG step, a singular value spectrum results from a singular value decomposition of the reshaped tensor, e.g. $A_{(ij)(kl)}$.  Therefore, the von Neumann entropy of the singular value spectrum at a particular TRG step corresponds to the entanglement entropy of a contiguous subblock of the  boundary (1+1)d quantum system at that step.  This is also the case for HOTRG \cite{ueda2014doubling}.

Near criticality the finite value of $\chi$ results in a crossover in the behavior of the entanglement entropy growth from linear to constant in TRG step (see Fig. (\ref{fig_TRG_vNE}) for an example); this is qualitatively like the FSS to FES crossover known for periodic MPS \cite{pirvu2012matrix} and also the scaling crossover near criticality in CTMRG \cite{nishino1996numerical}.  Further, Ueda et al. \cite{ueda2014doubling} \textit{quantitatively} confirmed the presence of FES in HOTRG flows of the critical partition function of the Ising model in the region after the crossover.  We therefore make the following conjecture: FES is generically valid after the von Neumann entropy crossover in TRG flows of critical partition functions.  Combining this conjecture with the behavior of entanglement in the FES regime of (1+1)d quantum systems (i.e., that the entanglement entropy is maximal very close to the true critical point) and the quantum-classical correspondence, we arrive at the simple idea behind the method described in the next section: the maximum of the TRG von Neumann entropy after the crossover gives a good approximation for the location of the phase boundary.  The benchmarks below validate this idea.  

In passing, we note here the result in Ref.~\onlinecite{pirvu2012matrix} that ground states of critical MPS rings in the FES regime correctly capture local universal properties in spite of having vanishing overlap with the true ground states.

\begin{table*}[t]
\caption{Numerical and theoretical phase transition temperatures ($k_\textrm{B} T/J$) for the $q$-state Potts models on the square lattice.  $\xi$ is the theoretically known correlation length \cite{buddenoir1993correlation, iino2019detecting} at the transition (when approached from the higher temperature phase).  The numerical values are obtained with our TRG von Neumann entropy method to five decimal places with different values of the bond dimension $\chi$.  There is good agreement with the theoretically known values.  Parameter sweeps are very cheap with our method: a single TRG flow with $\chi=40$ for q=4 at the (numerical) transition temperature 0.91083 takes about 30 seconds on a current desktop computer.}
\label{table_Potts}
\begin{tabular}{|c|c|c|c|c|c|}
\hline
q-state Potts & $q=3$ & $q=4$ & $q=5$ & $q=6$ & $q=10$ \\
sq. lattice	            & ($\xi=\infty$) & ($\xi=\infty$) & ($\xi=2512.2$) & ($\xi=158.9$)  & ($\xi=10.6$) \\ \hline
TRG, $\chi$=20   & 0.99795 & 0.90948 & 0.85037 & 0.80901  &  0.70553 \\ \hline
TRG, $\chi$=30   & 0.99453 & 0.91140 & 0.85099 & 0.80657  &  0.70257 \\ \hline
TRG, $\chi$=40   & 0.99494 & 0.91083 & 0.85193 & 0.80716  &  0.70247 \\ \hline
theory		   & 0.99497 & 0.91024 & 0.85153 & 0.80761  & 0.70123 \\ \hline
\end{tabular}

\end{table*}

\begin{table*}[t]
\caption{Numerical phase transition temperatures ($k_\textrm{B}T/|J_1|$) for the $J_1-J_2$ Ising model on the square lattice as determined with our TRG von Neumann entropy method to three decimal places and with Monte Carlo in Ref.~\onlinecite{murtazaev2015critical}, except the MC value for $\frac{J_2}{|J_1|}=0.6$ is approximated from the plot in Fig. (3) of Ref.~\onlinecite{kalz2008phase}.  Our method is very efficient (e.g. a single TRG flow with $\chi=30$ at the (numerical) transition temperature for $\frac{J_2}{|J_1|}=0.8$ completes in about 5 seconds on a current desktop computer), but still gives results in agreement with Monte Carlo.}
\label{table_J1J2}
\begin{tabular}{|c|c|c|c|c|c|}
\hline
$J_1$-$J_2$ sq. lattice & $\frac{J_2}{|J_1|}=0.1$     & $\frac{J_2}{|J_1|}=0.3$     & $\frac{J_2}{|J_1|}=0.4$     & $\frac{J_2}{|J_1|}=0.6$ & $\frac{J_2}{|J_1|}=0.8$  \\ \hline
TRG, $\chi$=20   & 1.946       & 1.256           &  0.868       &   0.973     & 1.568       \\ \hline
TRG, $\chi$=30   & 1.943       &  1.255          &    0.868        &   0.972     &  1.568   	\\ \hline
TRG, $\chi$=40   & 1.944       & 1.255           &    0.867        & 0.972     &   1.568	\\ \hline
Monte Carlo    & 1.952          &  1.258       &  0.873     & $\sim$0.95        &  1.567  	         \\ \hline
\end{tabular}

\end{table*}

\section{Method}
\begin{figure}
\includegraphics[scale=0.5]{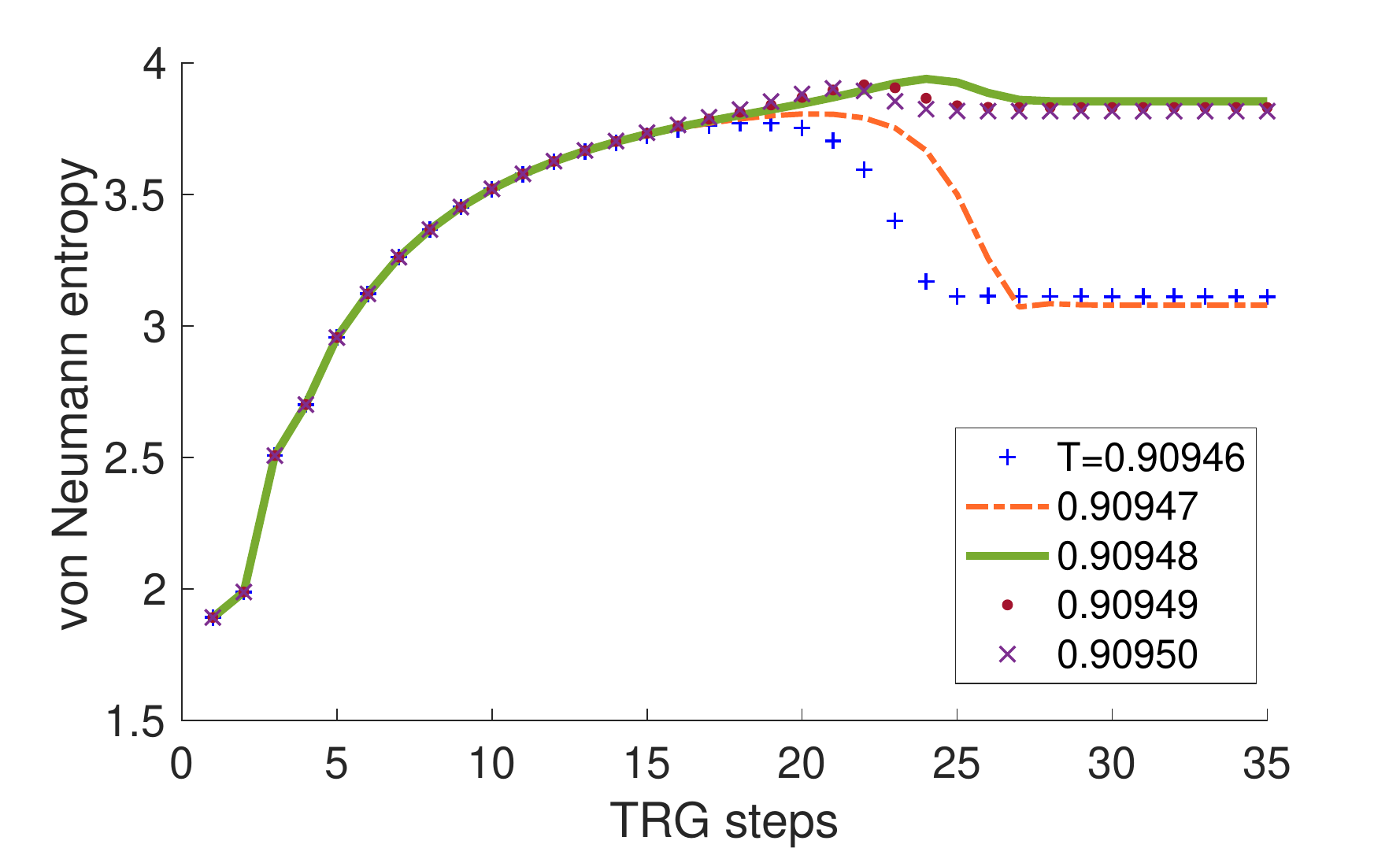}
\caption{von Neumann entropy of the singular value spectrum along TRG flows of the partition function at different temperatures near criticality for the $q=4$ Potts model on the square lattice.  The bond dimension chosen here is $\chi=20$, and the entropy peak along each of these flows is found in less than 1 second on a current desktop computer.  For a given $\chi$, the temperature that maximizes the entropy peak (in this case $T=0.90948$, solid line) is designated as the transition temperature.}
\label{fig_TRG_vNE}
\end{figure}

In our simulations we always find that the von Neumann entropy of singular values in TRG flows of partition functions reaches a maximum along the flow just before plateauing in the FES regime (see Fig.(\ref{fig_TRG_vNE}) for an example). Though larger lattice sizes (i.e. more TRG steps) intuitively give better results in the absence of truncation error, it is of no benefit to grow the lattice further once the FES regime is reached since the numerical correlation length has then reached the limit set by $\chi$.  The number of TRG steps $N$ after which the FES regime is entered is model- and $\chi$-dependent, but the generic existence of the entropy peak along TRG flows allows us to accommodate all scenarios in an algorithmically simple way.  Therefore, we implement our method around the peak value of the von Neumann entropy along individual TRG flows: for the TRG flow at a given temperature and $\chi$ we simply monitor the von Neumann entropy along the flow and find the step $N_\textrm{peak}$ after which the von Neumann entropy decreases for five consecutive steps.  We then record the von Neumann entropy at step $N_\textrm{peak}$ as the peak entropy value for that temperature and $\chi$.  As illustrated in Fig. (\ref{fig_TRG_vNE}), for a given $\chi$ the temperature that maximizes this peak entropy is designated as the transition temperature.  For first order transitions there is no emergent criticality or FES regime, but we may still use the same method by assuming that the physical correlation length is maximal at the phase boundary.  We show with benchmarks below that this method works very well with only modest $\chi$ for continuous, weakly first order, and regular first order phase transitions, and that it works for both unfrustrated and frustrated systems.

\section{Benchmarks: Potts models}  Here we benchmark our method with theoretical results for the $q$-state Potts models on the square lattice \cite{wu1982potts}; the Hamiltonian is
\begin{equation}
\mathcal{H}=-J\sum_{\langle i,j\rangle}\delta_{\sigma_i,\sigma_j},
\end{equation}
where $J>0$, $\delta$ is the Kronecker delta function, $\langle ~ \rangle$ denotes nearest neighbors, and $\sigma = 1,2,...,q$.  For $q\leq4$ the phase transition is theoretically known as continuous, and for $q>4$ it is first-order.  For $q=5$ the phase transition is very weakly first order (i.e. has a finite but very large correlation length); the strength of the first order nature increases with $q$ (i.e. the correlation length becomes smaller).

\begin{figure*}[t]
\includegraphics[scale=0.5]{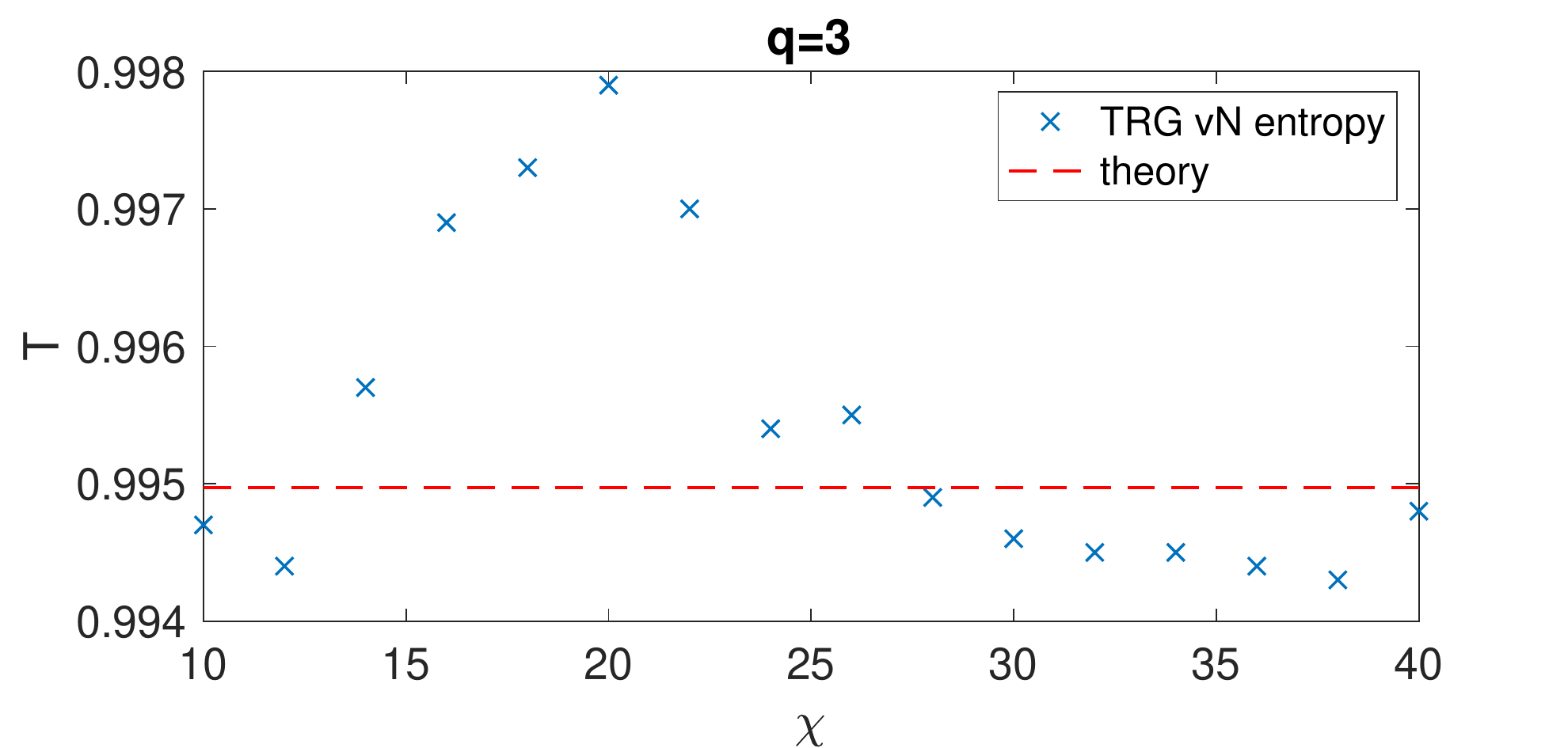}

\includegraphics[scale=0.5]{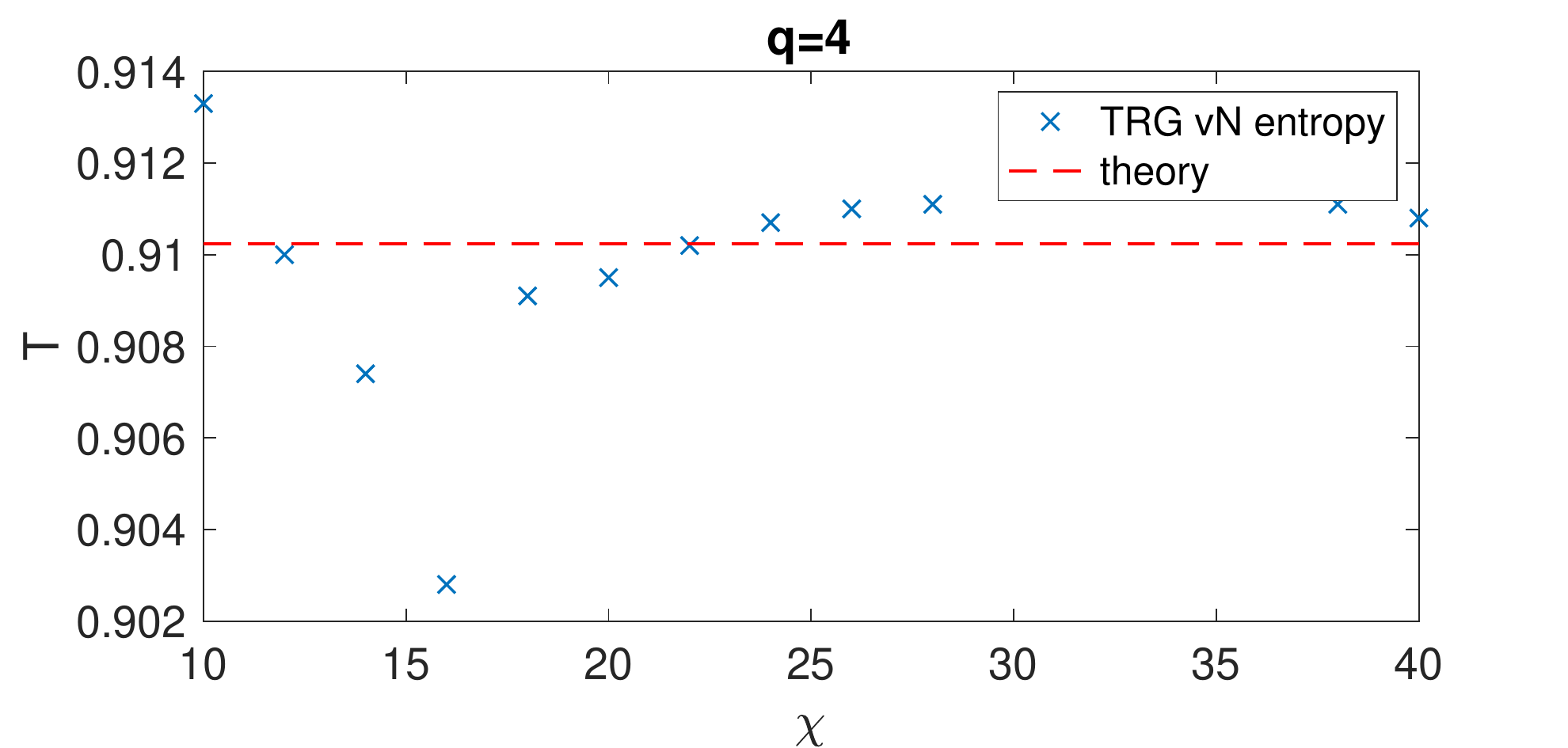}

\includegraphics[scale=0.5]{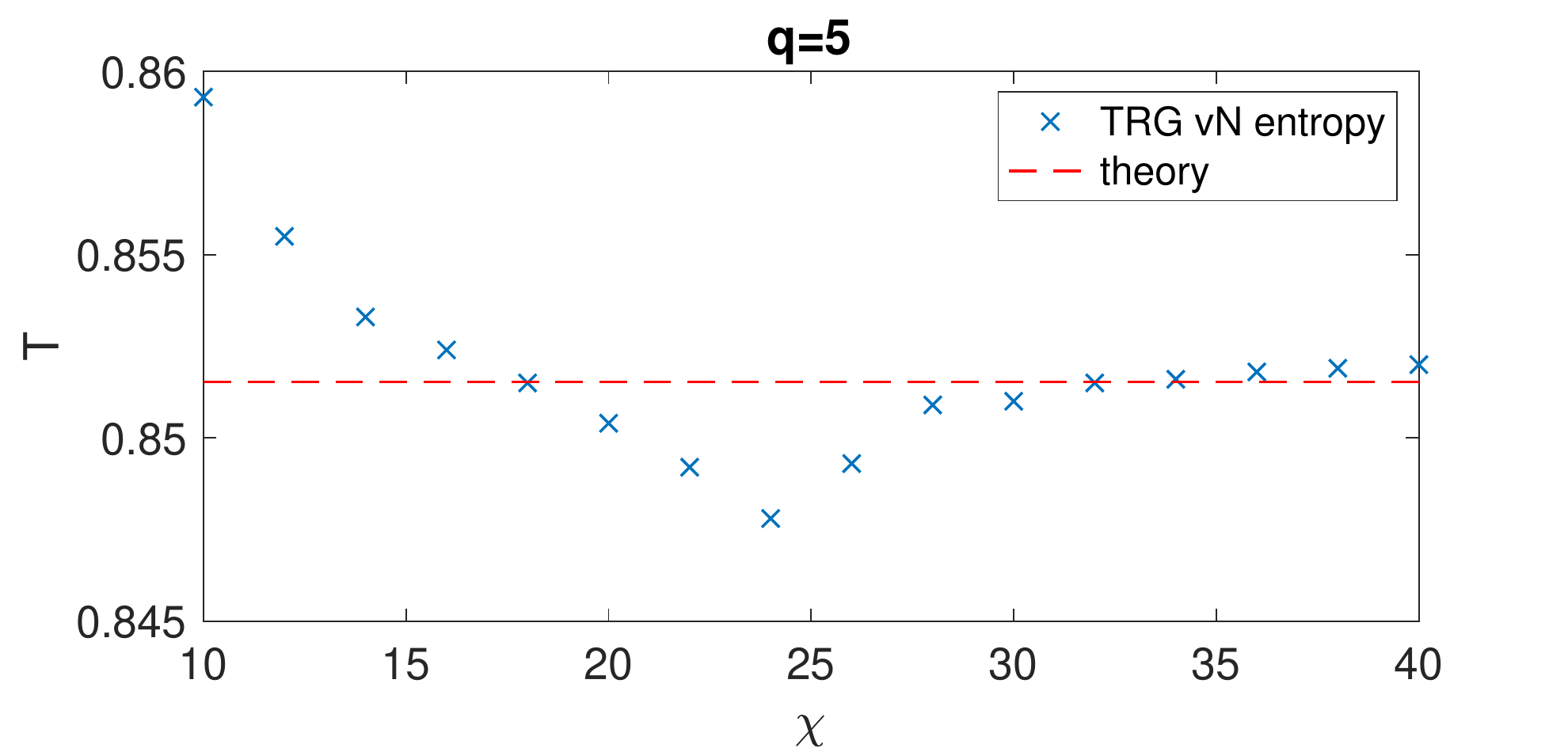}

\includegraphics[scale=0.5]{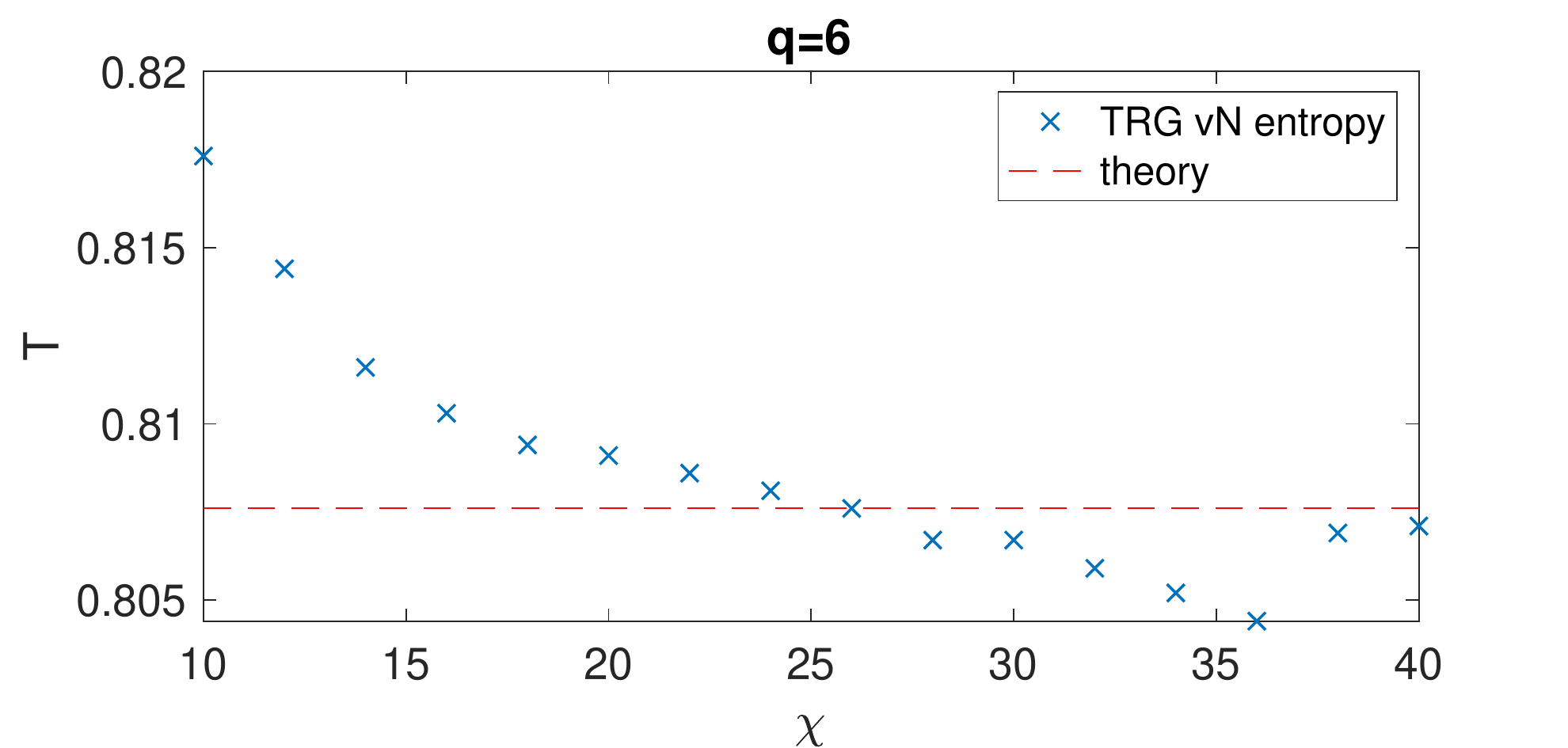}

\includegraphics[scale=0.5]{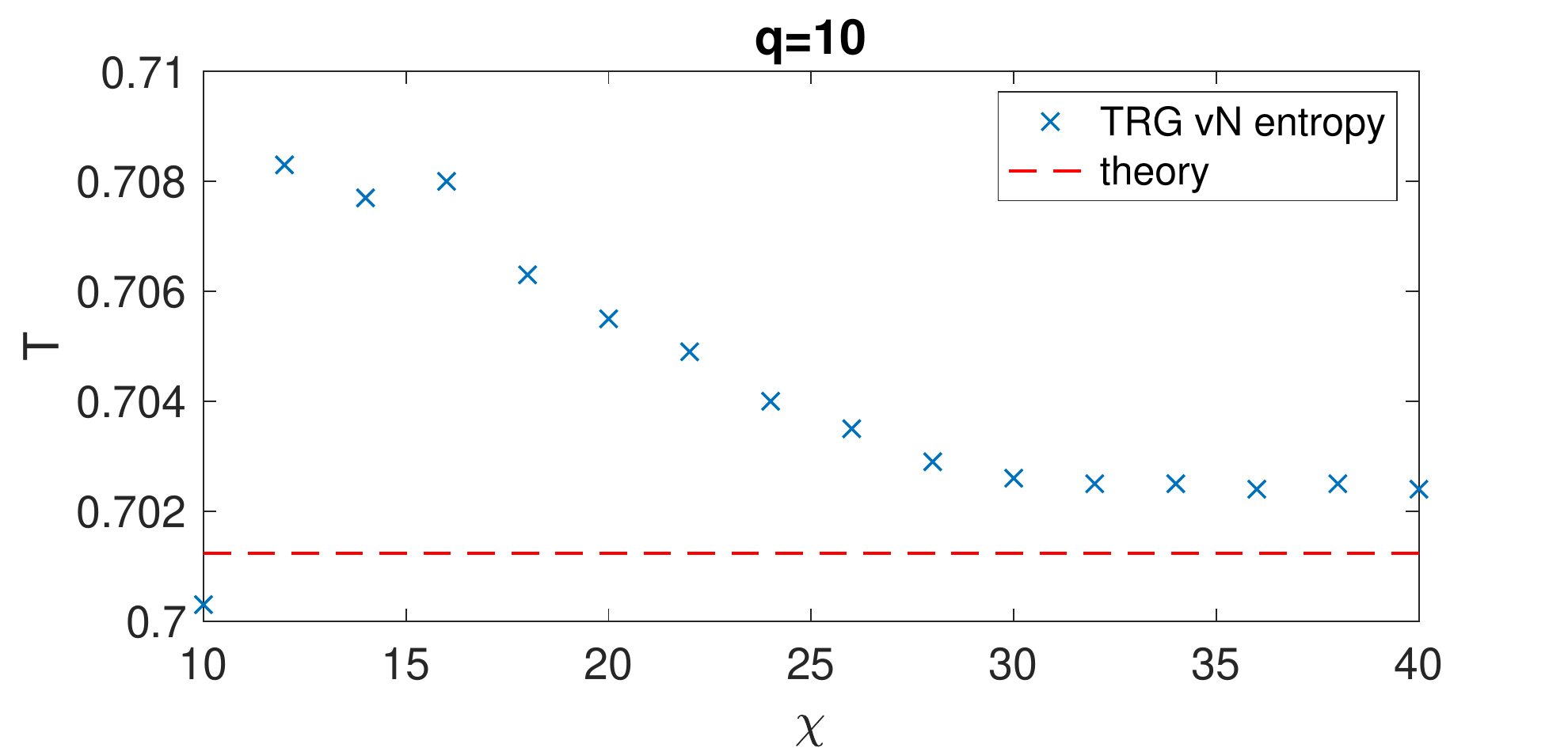}
\caption{Transition temperatures (with $k_\textrm{B}=|J_1|=1$) for the q-state Potts models on the square lattice from theory and as computed with our TRG von Neumann entropy method at different bond dimensions ($\chi$) to precision $10^{-4}$.}
\label{fig_Potts}
\end{figure*}

In Table \ref{table_Potts} we compare the transition temperatures ($k_\textrm{B}T/J$) from our method (to precision $10^{-5}$) with the theoretically known values $k_\textrm{B}T^{*}/J=1/\textrm{ln}(1+\sqrt{q})$. Further data at more values of $\chi$ is shown in Fig. (\ref{fig_Potts}).  Our method performs very well with only moderate values of $\chi$, and the numerical results (nonmonotonically) approach the theoretical results as $\chi$ increases.

\section{Application: J1-J2 Ising model} Here we compare our method's results with previous Monte Carlo results for the (frustrated) $J_1-J_2$ Ising model on the square lattice.  The Hamiltonian is

\begin{equation}
\mathcal{H}=J_{1}\sum_{\langle i,j\rangle}\sigma_i\sigma_j + J_{2} \sum_{\langle\langle i,j \rangle\rangle} \sigma_i \sigma_j , 
\label{eqn_J1J2}
\end{equation}
where $\sigma=\uparrow,\downarrow$, $\langle ~ \rangle$ denotes nearest neighbors, and $\langle\langle~\rangle\rangle$ denotes next nearest (i.e. diagonal) neighbors.

\begin{figure*}[t]
\includegraphics[scale=0.5]{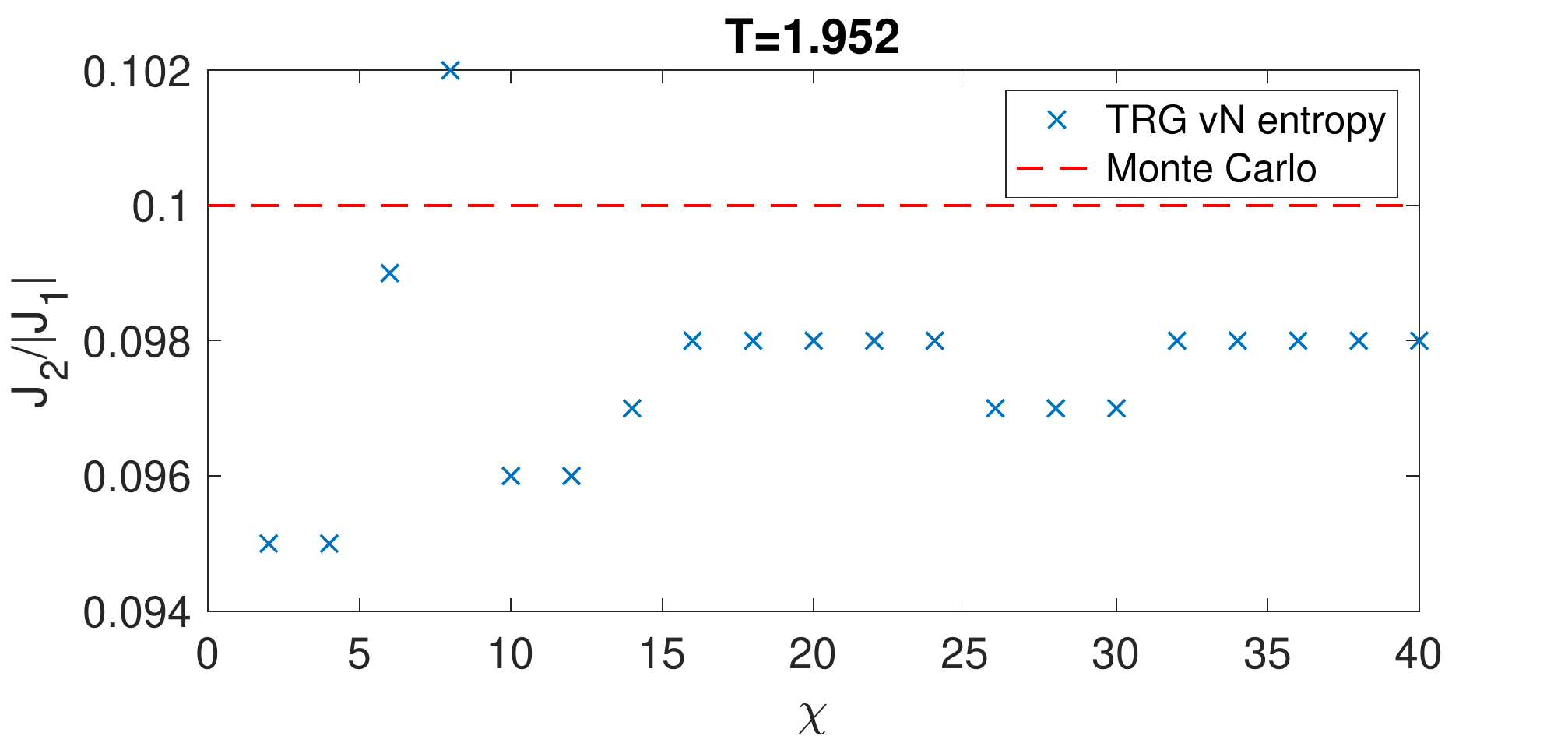}

\includegraphics[scale=0.5]{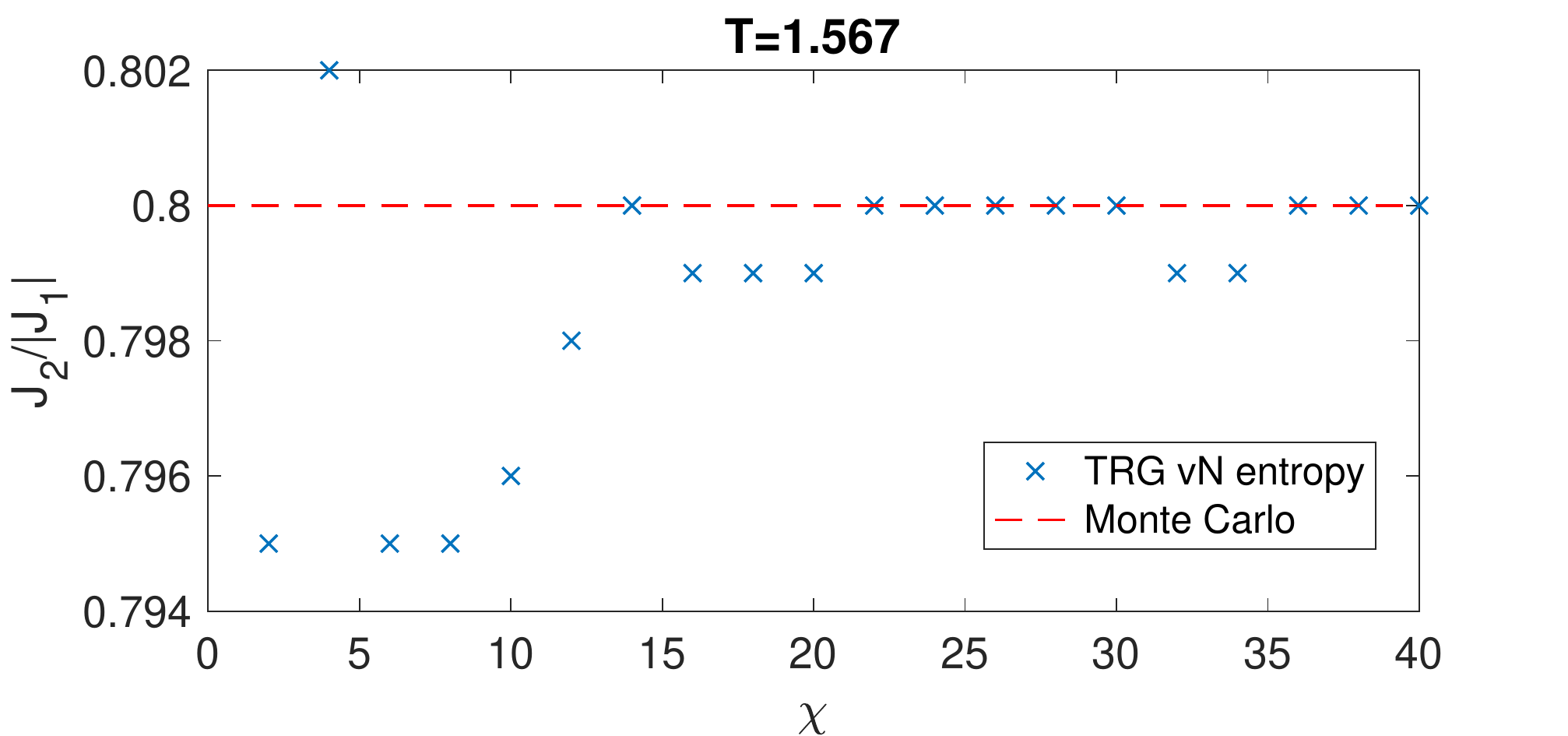}

\includegraphics[scale=0.5]{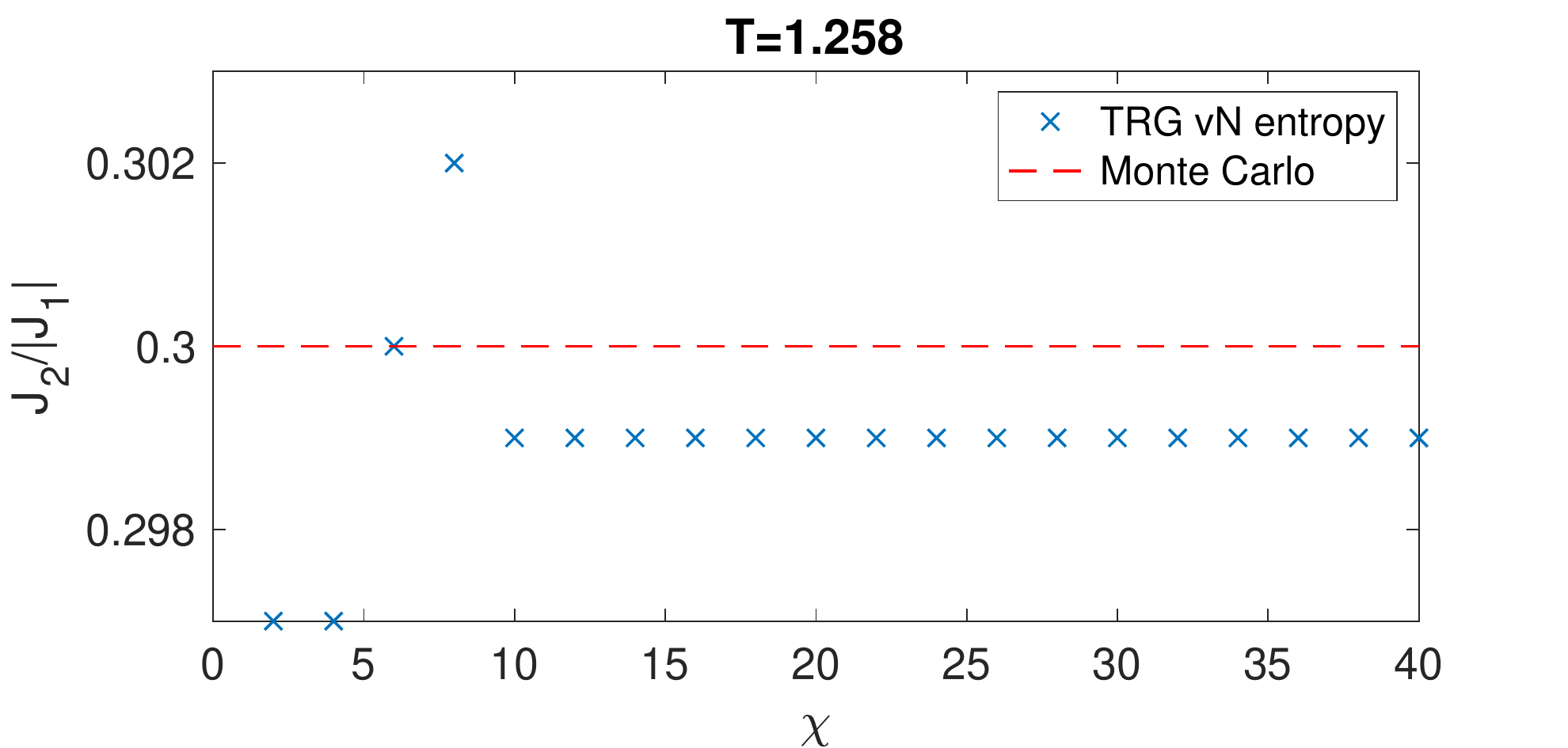}

\includegraphics[scale=0.5]{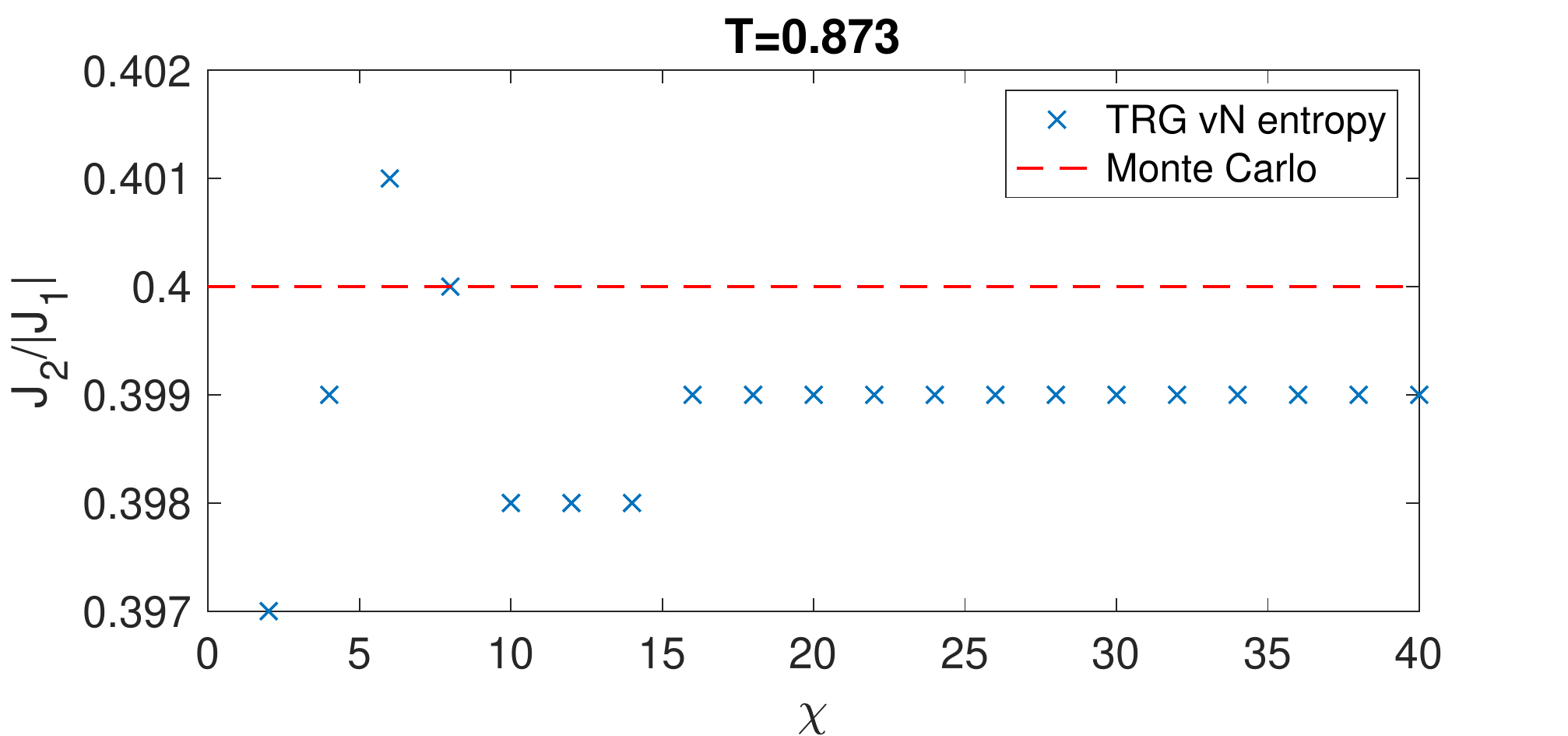}

\includegraphics[scale=0.5]{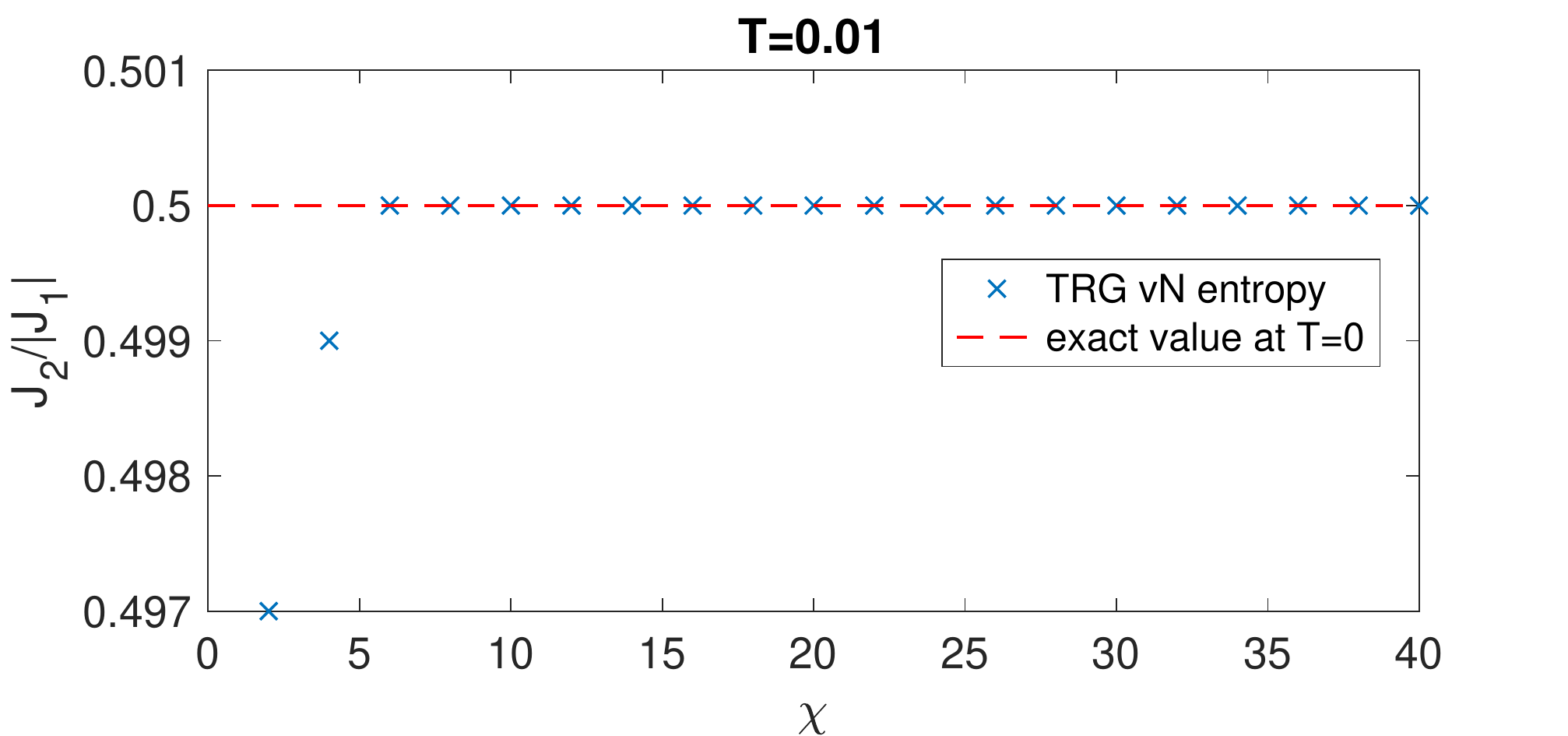}
\caption{Transition values of $J_2/|J_1|$ for the frustrated Ising model phase transitions denoted in Fig.(\ref{fig_J1J2phases}) at different $T$ (with $k_\textrm{B}=|J_1|=1$) as computed with our TRG von Neumann entropy method at different bond dimensions ($\chi$) to precision $10^{-3}$ and as computed with Monte Carlo in Ref. \onlinecite{murtazaev2015critical} or theoretically known.}
\label{fig_J1J2}
\end{figure*}

\begin{figure}
\includegraphics[scale=0.4]{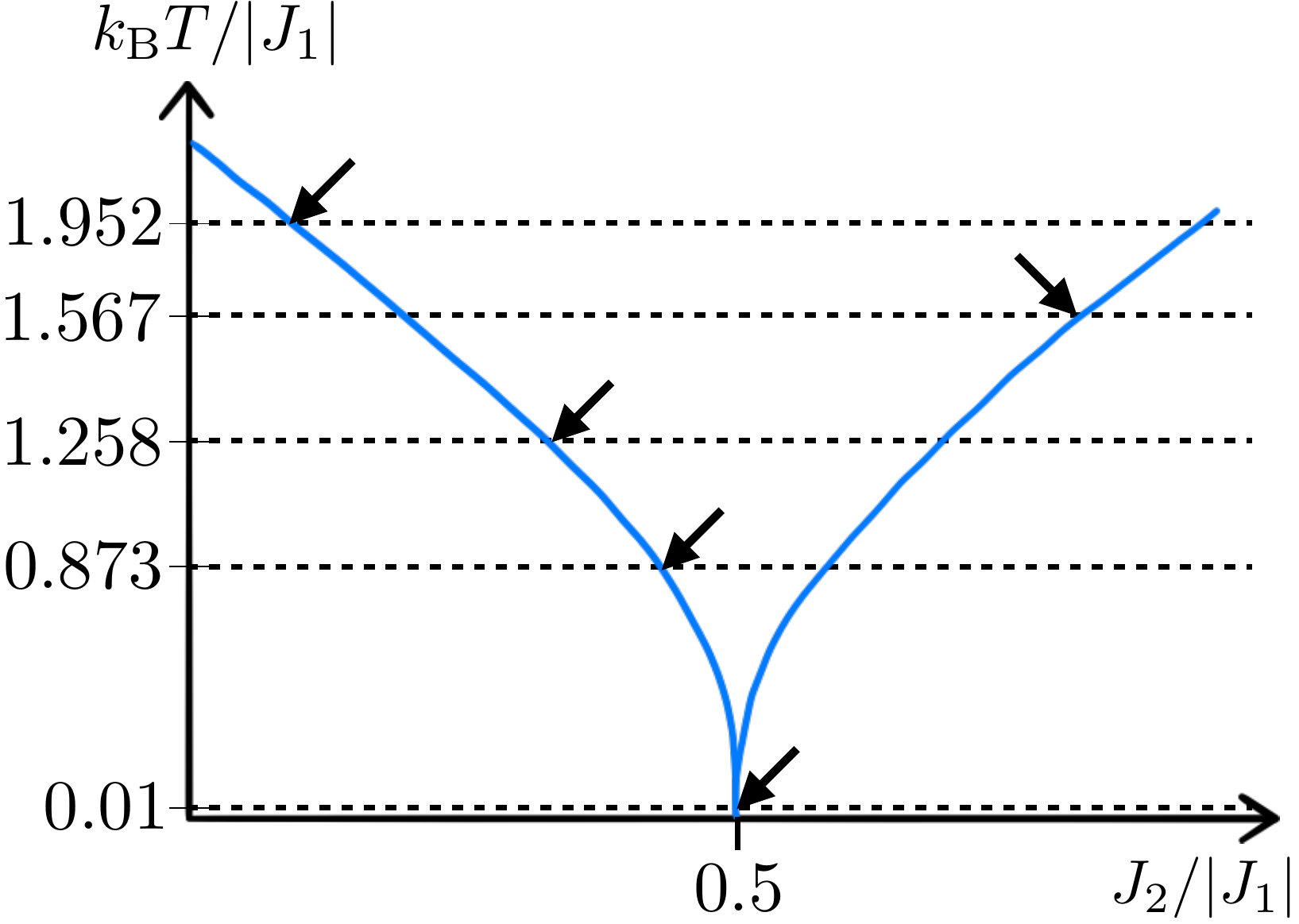}
\caption{Schematic diagram of the phase boundaries (blue solid lines) in the $J_1-J_2$ Ising model on the square lattice.  Arrows indicate the phase transitions analyzed in Fig. (\ref{fig_J1J2}) at the respective temperatures (dashed black lines).}
\label{fig_J1J2phases}
\end{figure}

In Table \ref{table_J1J2} we compare the transition temperatures from our method with the values obtained in the Monte Carlo studies in Ref.~\onlinecite{kalz2008phase,murtazaev2015critical}.  With only moderate values of $\chi$, the results from our method match closely with the Monte Carlo results.  Further comparison between the methods is provided in Fig. (\ref{fig_J1J2}), which displays the computed transition values of $J_2/|J_1|$ at fixed temperatures with different values of $\chi$.  As illustrated in the schematic phase diagram in Fig. (\ref{fig_J1J2phases}), this model has more than one phase transition in $J_2/|J_1|$ over a range of temperatures; we arbitrarily choose one at each temperature for the data in Fig. (\ref{fig_J1J2}).

\section{Summary}
By leveraging the (1+1)d quantum to 2d classical correspondence known to exist in TRG and what is already known about the behavior of the entanglement entropy in MPSs near criticality, we have shown that the von Neumann entropy of the singular values that arise in computationally efficient TRG flows of partition functions near first-order, weakly first-order, and continuous phase transitions can provide an accurate location of phase transitions in spite of the presence of large truncation errors.  

Due to it's combination of simplicity, efficiency, and accuracy, the method presented here has the potential to become a standard tool for locating phase boundaries of 2d classical lattice models.  Though restricted to only phase boundary location of two-dimensional classical systems, the method presented here is extremely fast, much simpler than performing thermodynamic analysis, does not require encoding of symmetries, and is applicable to both first-order and continuous phase transitions.

We note that the method described here can also work with two-dimensional HOTRG instead of TRG; it is an avenue for further investigation to see if using this method with HOTRG instead of TRG can yield better accuracy at similar cost.

\section{Acknowledgements}
AAG acknowledges discussions with Kai-Hsin Wu and Glen Evenbly.  This work is supported by the MOST in Taiwan through Grants No. 107-2112-M-002 -016 -MY3, and 105-2112-M-002 -023 -MY3. 

\appendix*
\section{}
Tensor network representations of classical partition functions are not unique \cite{zhao2010renormalization}.  We detail here the particular representation that we use for the TRG flows of the partition function of the frustrated Ising model of Eq. (\ref{eqn_J1J2}).  The representation is that of a single repeated tensor, whose construction was pointed out by Evenbly \cite{evenbly}.

\begin{figure*}
\includegraphics[scale=0.46]{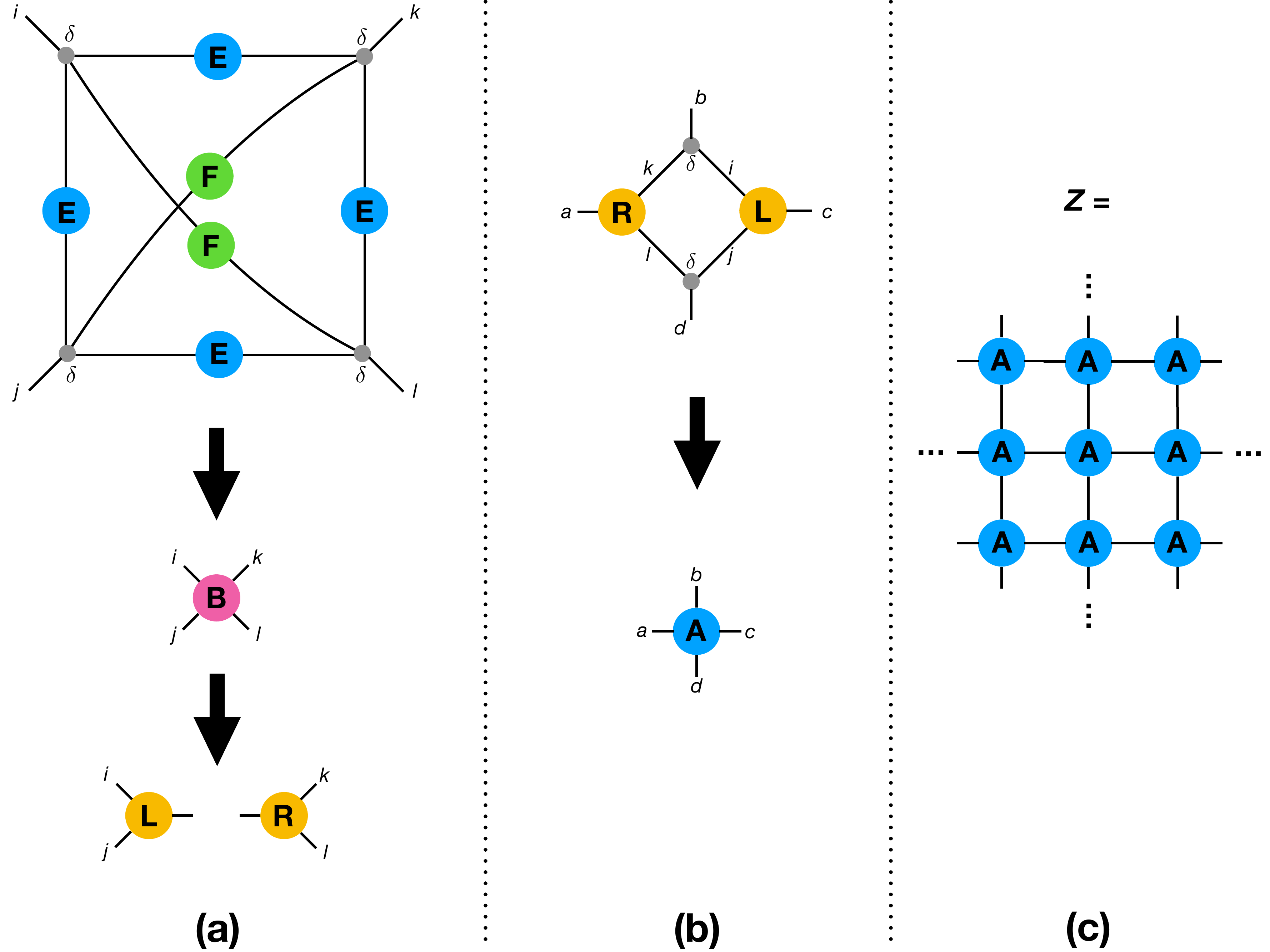}
\caption{Illustration of the construction of the repeated single tensor (A) that represents the partition function $Z$ for the frustrated Ising model in Eq. (\ref{eqn_J1J2}). The tensor A is used as the initial tensor for the TRG flows.  The definitions of the tensors is given in the main text.  (a) An initial network of tensors is contracted then split.  (b) The split parts are then contracted with kronecker delta functions to form the initial tensor A.  (c) The partition function can be represented as a contraction of the network of A tensors.}
\label{fig_appendix}
\end{figure*}

We illustrate the construction in Fig.(\ref{fig_appendix}). The strategy is to contract a single plaquette of the partition function on the real lattice and then split and reshape it so that the next nearest neighbor interactions of the original plaquette become nearest neighbor interactions between the new tensors.  Let E and F be $2\times2$ symmetric matrices with elements $E_{mn}=\delta_{mn} e^{-\beta J_1/2} + (1-\delta_{mn}) e^{\beta J_1/2}$ and $F_{mn}=\delta_{mn} e^{-\beta J_2} + (1-\delta_{mn}) e^{\beta J_2}$, where $\delta$ is the kronecker delta function.  Matrices E and F along with 4-index kronecker delta functions form the tensor network at the top of Fig. (\ref{fig_appendix}a), where each leg represents an index of the associated tensor and connected legs represent a contraction of the corresponding tensors over the corresponding indices. The network is first contracted into tensor $B_{ijkl}$ and then reshaped into the symmetric matrix $B_{(ij)(kl)}$ and split with an eigendecomposition into matrices $L_{(ij)x}$ and $R_{x(kl)}$ such that $B_{(ij)(kl)}=L_{(ij)x}R_{x(kl)}$, where $L_{(ij)x}=U_{(ij)v}\textrm{sign}(D)_{vw}(\sqrt{|D|})_{wx}$ and $R_{x(kl)}=(\sqrt{|D|})_{xy}(U^{-1})_{y(kl)}$ (Einstein summation convention applies).  $L$ and $R$ are then reshaped into three-index tensors and contracted with 3-index kronecker delta functions to yield the single repeated tensor for the partition function: $A_{abcd}=R_{akl}\delta_{kbi}L_{ijc}\delta_{ldj}$.

\bibliographystyle{apsrev4-1}
\bibliography{Ref}

\end{document}